\documentclass[11pt]{article} 
\usepackage{fixltx2e,fix-cm}
\usepackage{amssymb}
\usepackage{amsmath}
\usepackage{graphicx}
\usepackage{subfigure}
\usepackage{makeidx}
\usepackage{multicol}
\graphicspath{{plots/}}
\usepackage{hyperref}
\usepackage{fullpage}
\usepackage{bm}
\usepackage{graphicx,psfrag,epsf}
\usepackage{natbib}
\usepackage{xcolor}

\hypersetup{
    colorlinks=true,
    linkcolor=blue, 
    urlcolor=black,
    citecolor=blue, 
    }

\title{Nonstationary Gaussian Process Surrogates}
\author{Annie S.~Booth\thanks{Corresponding author: Department of Statistics, 
	Virginia Tech, {\tt anniees@vt.edu}} 
	\and Andrew Cooper\thanks{Department of Statistics, Virginia Tech} 
	\and Robert B. Gramacy\footnotemark[2]}
\date{\today}
\begin{document}

\maketitle 
\bigskip

\begin{abstract}
We provide a survey of nonstationary surrogate models which utilize
Gaussian processes (GPs) or variations thereof, including
nonstationary kernel adaptations, partition and local GPs, and
spatial warpings through deep Gaussian processes.  We also overview
publicly available software implementations and conclude with
a bake-off involving an 8-dimensional satellite drag computer 
experiment.  Code for this example is provided in a public git 
repository.
\end{abstract}

\section{Introduction}

Computer simulations are becoming increasingly relevant as methods of data
collection, whether to augment or stand in place of actual physical
observations, particularly when physical observations are too costly or
impossible to obtain.  Such computer ``experiments'' pervade many scientific
fields including aerospace engineering
\citep[e.g.,][]{renganathan2021enhanced}, cosmology
\citep[e.g.,][]{moran2022mira}, epidemiology
\citep[e.g.,][]{andrianakis2015bayesian}, biology
\citep[e.g.,][]{johnson2008microcology}, and physiology
\citep[e.g.,][]{plumlee2016calibrating}.  Computer experiments aim to mimic the
real-world.  As they have been refined over the years, and their fidelity
improved, they have increasingly required more computation. 
This limits the number of simulator evaluations that may be obtained, even with modern
supercomputing resources.  In such situations, a statistical ``surrogate'' model
(also called an ``emulator''), which is trained on a limited amount of
simulation data, can be useful as a stand-in for additional
expensive simulations.   That is, as long as it can provide accurate
predictions with appropriate uncertainty quantifiaction (UQ) at untried input
configurations. Surrogates can be a crucial tool for turning limited computer
simulation data into actionable scientific conclusions.

The stipulation we made above bears repeating; the usefulness of a surrogate
relies on its ability to provide appropriate {\it uncertainty quantification}.
UQ is a priority because surrogate models are most often tools to be used to
another end: objectives we refer to as ``downstream tasks.''  For example,
the objective may be to maximize the response, i.e., global optimization.
Consider a simulation of a wind turbine in which input configurations include
turbine height, angle, and blade dimensions.  These factors affect the energy output
of the turbine \citep[e.g.,][]{marten2013qblade}. A natural research objective
is to find the configuration that maximizes energy output.  Adaptive
surrogate-based design targeting a global optimum is commonly
termed Bayesian optimization
\citep[BO;][]{jones1998efficient,eriksson2019scalable,binois2022survey}.
[Often, BO is used in machine learning to optimize settings of model
hyperparameters.]  In this endeavour, knowing where high outputs are found is
just as important as identifying input regions of high uncertainty to explore
further (striking a balance between exploitation and exploration).  Another
objective may be to ensure a simulated system meets certain safety
thresholds by quantifying the probability of a system failure
\citep[e.g.,][]{booth2024hybrid,booth2023contour}.  Consider a
simulation of an aircraft whose response variable is the amount of aircraft
vibration.  It is dangerous for the vibrations to exceed certain safety
levels; thus, the design objective is to identify the input regions that result
in unsafe vibration conditions \citep[e.g.,][]{stanford2022gradient} so they
can be avoided in practice.  Both these objectives fall under the umbrella of
``sequential design'' or ``active learning,'' in which input configurations are
chosen sequentially and strategically to target specific learning outcomes
based on surrogate information.  Another common objective is simulator
calibration, in which tuning parameters of the simulator are calibrated to
physical observations \citep{kennedy2001bayesian}.  Sequential
design has also been entertained in calibration contexts
\citep[e.g.,][]{koermer2023active}.

In summation, we need a statistical model that can handle the nonlinear
response surfaces typical of computer simulations, work with limited training
data, and provide accurate predictions with thorough UQ to facilitate
downstream tasks. The ``go-to'' surrogate model is a Gaussian process (GP).
GPs are nonlinear, nonparametric regression models that are preferred for
their predictive prowess; see \cite{santner2018design} and
\cite{gramacy2020surrogates} for reviews of GPs for surrogate modeling.  While
GPs are the canonical surrogate modeling choice, the computer experiment
community is only one of three communities where GPs play a vital role.  GPs
are widely used in both spatial statistics
\citep{banerjee2003hierarchical,stein1999interpolation, cressie2015statistics}
and machine learning \citep[ML;][]{rasmussen2005gaussian}.  Although GPs unite
these communities, there are subtle differences in applications which,
naturally, inform modeling decisions. In spatial statistics, where GP
regression is known as kriging, the focus is on low dimensional input spaces,
often with missing data and anticipated measurement error.  The norm in ML is
high input dimension, large data sizes, and lots of noise.  On the contrary,
computer experiments commonly live in modest dimensional realms with
little-to-no noise \citep[though there are exceptions,
e.g.,][]{baker2022analyzing}, and small-to-moderate data sizes due to large
computational costs. Nevertheless, many of the key innovations for GP
surrogate modeling arise from the spatial and ML communities.  We will revisit
this motif throughout this chapter.

Despite their nonlinear flexibility, traditional GPs are limited by the
assumption of {\it stationarity}.  Stationary GPs rely solely on
pairwise distances between inputs, so they must impart similar dynamics across
the entire input space (more on this in Section \ref{sec:gp}). Yet
nonstationary response surfaces that exhibit regime shifts, sharp turns,
and/or space-varying dynamics are common in computer experiments. For example,
consider an aerospace simulation of a rocket re-entering the atmosphere
\citep{pamadi2004aerodynamic}, as featured in \citet[][Chapter
2]{gramacy2020surrogates}.  A visual is provided later in Figure
\ref{fig:tree}.  There is an abrupt shift in aeronautical dynamics when speeds
cross the sound barrier.  A stationary GP is unable to accommodate the stark
ridge in the response surface at this boundary. Alas, we are finally at the
motivation behind this chapter: advancements to typical GP models that allow
nonstationary flexibility without sacrificing predictive prowess or
uncertainty quantification.

There has been much work on nonstationary GP modeling, with many
contributions from spatial and ML communities.  We see these methods falling
into three categories:
\begin{itemize}
	\item {\bf Nonstationary kernels:}  adapt
	the GP kernel formulation spatially so that dynamics are no longer strictly a
	function of relative distance.  These methods originate from spatial
	statistics, where the focus is on low input dimension.
	\item {\bf Divide-and-conquer:}  partition
	the input space and use traditional stationary GPs independently
	in each region.
	Localization is common in surrogate modeling and in geospatial contexts,
	but such schemes forfeit a degree of global scope when strong long-range
	dynamics are present.
	\item {\bf Spatial warping:}  nonlinearly map inputs so that the process
	can be depicted as plausibly stationary.  The most recently
	popular of these these is the ``deep Gaussian process''
	\cite[DGP;][]{damianou2013deep} which uses a stationary GP as the
	warping function.  DGPs combine nonstationary and global modeling, albeit
	at the cost of some additional computation.  Although this modern approach
	is inspired by recent advances in deep neural networks (DNNs) in
	machine learning, it actually originated in the spatial and computer
	modeling literature a decade before DNNs became popular.
\end{itemize}
In the remainder of this chapter, after a brief review of stationary GPs, we 
will address each of these three categories, keeping an eye on surrogate modeling 
applications and publicly available implementations.

\section{Gaussian process fundamentals}\label{sec:gp}

Allow us to introduce GPs in their simplest form, to motivate the
upgraded versions we will discuss later.  Let 
$f: \mathbb{R}^d \rightarrow \mathbb{R}$ represent a black-box
computer simulation.  Denote a $d$-dimensional vectorized
input configuration as $\mathbf{x}$, with corresponding simulator output
$y = f(\mathbf{x})$.  Similarly, let $X_n$ denote the row-combined
matrix of $n$ input configurations with $\mathbf{y}_n = f(X_n)$. A GP prior
assumes a multivariate normal distribution over the response,
$\mathbf{y}_n \sim N\left(\boldsymbol\mu, \boldsymbol{\Sigma}(X_n)\right)$.  It is 
possible to model the prior mean as a linear combination of inputs, i.e.,
$\boldsymbol\mu = X_n\boldsymbol\beta$, but we will assume 
$\boldsymbol\mu = \mathbf{0}$, as is common after centering, without 
loss of generality.  The prior covariance matrix $\boldsymbol{\Sigma}(X_n)$
is an $n\times n$ matrix with elements 
$\boldsymbol{\Sigma}(X_n)^{ij} = \Sigma(\mathbf{x}_i, \mathbf{x}_j)$ denoting the 
covariance between the $i^{th}$ and $j^{th}$ input locations. 

A common choice for $\boldsymbol{\Sigma}(\cdot)$ is the Gaussian or squared exponential
kernel,
\begin{equation}\label{eq:kernel}
\Sigma(\mathbf{x}_i, \mathbf{x}_j) = \Sigma(||\mathbf{x}_i - \mathbf{x}_j||) = 
	\sigma^2 \left(\mathrm{exp}\left(- \frac{||\mathbf{x}_i - \mathbf{x}_j||^2}
	{\phi}\right) + \nu\mathbb{I}_{i=j}\right),
\end{equation}
where $\sigma^2$ acts as a scale parameter, $\phi$ is a lengthscale, 
and $\nu$ is a nugget/noise parameter.  These parameters may be estimated
through maximum likelihood estimation 
\citep[as in][]{gramacy2020surrogates} or sampled
through Markov Chain Monte Carlo
\citep[MCMC; as in][]{sauer2023active}.  In deterministic
computer simulations, we often fix $\nu$ at a small constant, on the scale
of $1\times 10^{-6}$.  One may instead use a Mat\'ern kernel 
\citep{stein1999interpolation}, parameterized similarly.  These 
covariance functions are {\it stationary} as they are solely functions
of relative distances.  Even if
we expand to separable vectorized lengthscales $\phi = [\phi_1, \dots, \phi_d]$
\citep[][Chapter 5]{gramacy2020surrogates},
the kernels are unable to encode information aside from (possibly scaled)
relative distances.  This forces similar dynamics across the entire domain,
i.e., {\it stationarity}.

Conditioned on observed training data, posterior predictions at 
$n_p$ input locations,  $X_p$, follow
\begin{equation}\label{eq:gppred}
\mathbf{y}_p \mid X_n, \mathbf{y}_n \sim N\left(\boldsymbol\mu^\star, 
\boldsymbol{\Sigma}^\star\right)
\quad
\textrm{where}
\quad
\begin{array}{l}
\boldsymbol\mu^\star = \boldsymbol{\Sigma}(X_p, X_n)
\boldsymbol{\Sigma}(X_n)^{-1}\mathbf{y}_n \\
\boldsymbol{\Sigma}^\star = \boldsymbol{\Sigma}(X_p) - 
\boldsymbol{\Sigma}(X_p, X_n)\boldsymbol{\Sigma}(X_n)^{-1}
\boldsymbol{\Sigma}(X_n, X_p), 
\end{array}
\end{equation}
and $\boldsymbol{\Sigma}(X_p, X_n)$ is the $n_p \times n$ matrix containing the covariances
between each row of $X_p$ and each row of $X_n$. 
These closed form analytic posterior moments are convenient, but they 
rely heavily on the choice of covariance kernel -- notice
how frequently $\boldsymbol{\Sigma}(\cdot)$ features in (\ref{eq:gppred}).  
If the kernel is an ill-fit for the response surface dynamics, GP
predictions will be handicapped.

\begin{figure}[h]
\centering
\includegraphics[width=16cm]{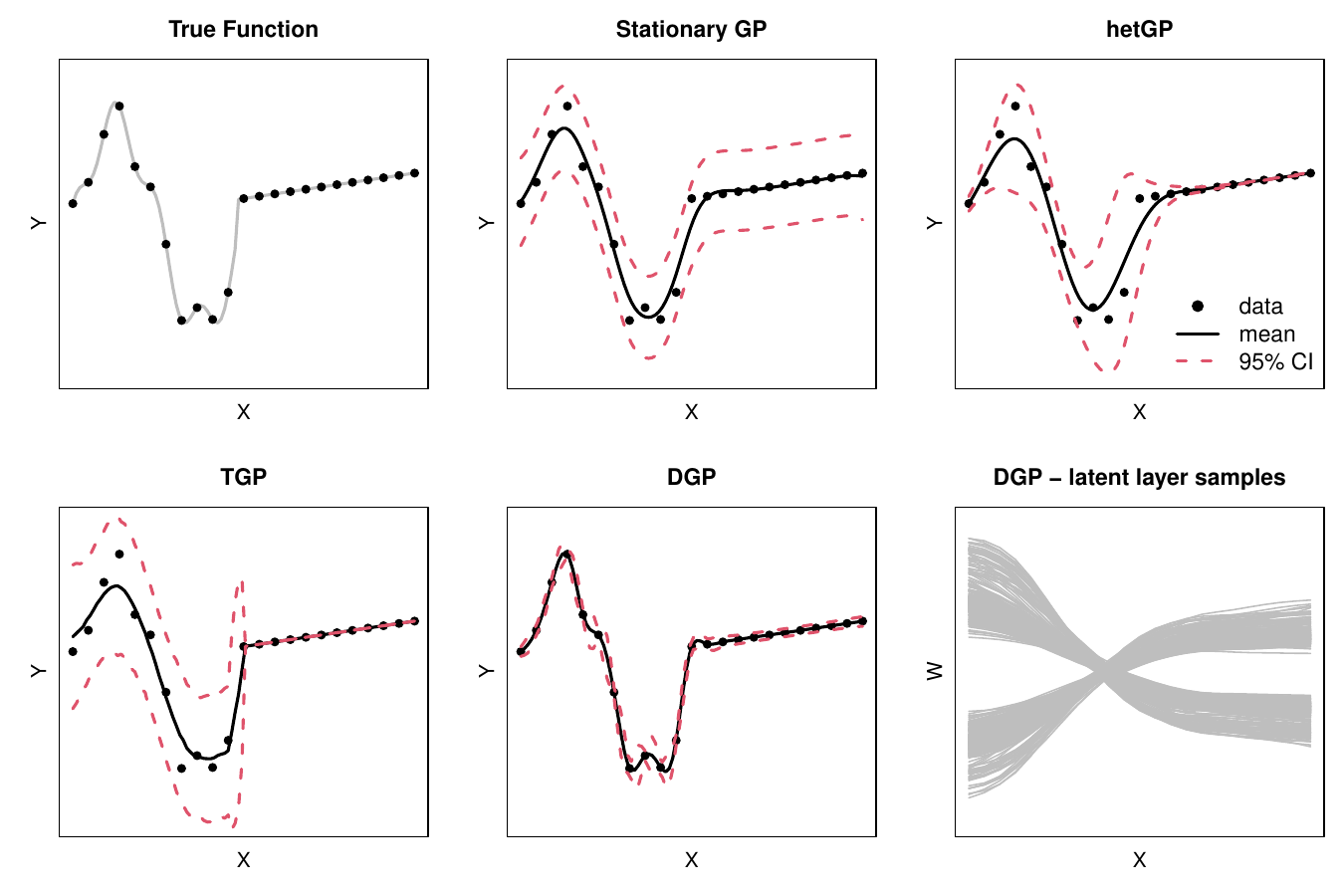}
\caption{Various surrogate fits to the piecewise Higdon function.
Training data is observed without noise, but each surrogate is tasked 
with learning the noise level.  The lower right panel shows elliptical
slice samples of the DGP's latent layer, stretching
inputs in the left region and compressing inputs in the right region.}
\label{fig:higdon}
\end{figure}

For example, consider the nonstationary ``Higdon function'' generated 
piecewise following \cite{gramacy2004parameter}, shown in the 
upper left panel of Figure \ref{fig:higdon}.  The left region is high signal, but
the right region is far less interesting.  Although the training data is observed
without noise, we allow surrogates to estimate a global noise parameter -- an additional
test of surrogate proficiency in this setting.  
A stationary GP fit is provided in the upper middle panel.  In trying
to reconcile the disparate regimes, it oversmooths in the left region and 
overinflates variance predictions in the right region.  We shall revisit
the other panels of the Figure in due course, after introducing recently popular
forms of nonstationary modeling.

\section{Nonstationary kernels}\label{sec:kernels}

If a stationary covariance kernel is not an appropriate fit, a natural
remedy is to change the kernel to one that is more flexible, while
still maintaining positive definiteness.  A nonstationary kernel 
$\boldsymbol{\Sigma}_\mathrm{ns}(\cdot)$ must rely on more than relative distances, i.e., 
$\boldsymbol{\Sigma}_\mathrm{ns}(\mathbf{x}_i, \mathbf{x}_j) \neq 
\boldsymbol{\Sigma}(||\mathbf{x}_i - \mathbf{x}_j||)$.  This advancement requires
introducing auxilary quantities into the covariance, which are typically unknown
and ideally learned from training data.  

\cite{higdon1999non} allowed for nonstationary kernels through
process convolution, i.e., $\boldsymbol{\Sigma}_\mathrm{ns}(\mathbf{x}_i, \mathbf{x}_j) 
= \int k_{\mathbf{x}_i}(u)k_{\mathbf{x}_j}(u)du$ where $k_{\mathbf{x}_i}$ 
is a squared exponential kernel centered at $\mathbf{x}_i$. 
They utilized a hierarchical Bayesian framework
and sampled unknown quantities through MCMC, allowing for full UQ.  
\cite{paciorek2003nonstationary}
later generalized this work to the class of Mat\'ern kernels,
with further development by \cite{katzfuss2013bayesian}.
The use of process convolutions for nonstationary GP modeling is popular
and has recent use cases \citep[e.g.,][]{nychka2018modeling,wang2020non}.
Other strategies involve adjustments to the underlying structure of the
problem itself.  \cite{bornn2012modeling} expand input dimensions to find a
projection into a reasonably stationary domain.  [There are some parallels here
to the the warping methods we introduce in Section \ref{sec:warp}.]
\cite{nychka2002multiresolution} opt to represent the covariance as a
linear combination of basis functions. 

Another way to incorporate flexibility into the kernel is by introducing
functional hyperparameters ($\sigma^2, \phi, \nu$ in (\ref{eq:kernel}), which
were previously assumed constant).  \cite{heinonen2016non} placed functional
Gaussian process priors on $\sigma^2(x)$, $\phi(x)$, and $\nu(x)$ and
entertained two Bayesian inferential methods: maximum posterior estimates and
full MCMC sampling.  \cite{binois2018practical} later focused on
heteroskedastic modeling of $\nu(x)$, for situations where the variance (as
opposed to the correlation or entire covariance structure) is changing in the
input space.  Their approach emphasizes computational speed through use of
Woodbury identities by taking advantage of replicates in the design. Yet,
there are pitfalls to introducing too much kernel flexibility.  Together, the
lengthscale $\phi$ and the nugget $\nu$ facilitate a signal-to-noise
trade-off.  In settings where the noise level is not yet pinned down, it is
impossible to distinguish between signal and noise.  In a computer surrogate
modeling framework such estimation risks can be mitigated by designing the
experiment carefully.  For example, in the context of heteroskedastic
modeling, one can learn where additional replicates are needed in order to
separate signal from noise \citep{binois2019replication}.

Off-the-shelf implementations of nonstationary kernels are
not readily available.  While there are some {\sf R} packages offering
nonstationary kernel implementations, such as {\tt convoSPAT} \citep{convoSPAT}
and {\tt BayesNSGP} \citep{BayesNSGP}, these are targeted towards spatial applications
and are only implemented for two-dimensional inputs. 
The heteroskedastic GP (hetGP) of \cite{binois2018practical}, however, is neatly wrapped 
in the {\tt hetGP} package for {\sf R} on CRAN \citep{hetGP} and is ready-to-use
on multi-dimensional problems.  We provide
a visual of a hetGP fit to the Higdon function in the upper right panel of Figure 
\ref{fig:higdon}, although we acknowledge that this example is a mismatch to the
hetGP functionality.  The hetGP model offers nonstationarity in the {\it variance},
but the Higdon function example exhibits nonstationarity in the {\it mean}.
Nevertheless, this benchmark still provides an interesting visual of the
flexibility of hetGP.  Although it oversmooths (its predicted mean matches that of the
stationary GP), it more properly allocates uncertainty in the linear region.

Perhaps limited software availability in this realm is a byproduct of the fact
that nonstationarity is wrapped up in the computational bottlenecks of
large-scale GPs.  Even with a flexible model, nonstationary dynamics will not
be revealed without enough training data in the right places.  In spatial
statistics, where nonstationary kernels have taken hold as the weapon of
choice, many of the methodological advances explicitly target both challenges
at once: enhancing modeling fidelity while making approximations to deal with
large training data
\citep[e.g.,][]{grasshoff2020scalable,huang2021nonstationary,noack2023exact}.
In surrogate modeling, divide-and-conquer methods (discussed next) can kill
those two birds with one stone.  When one has the ability to augment training
data at will -- say, through running new computer simulation -- data can be
acquired specifically to address model inadequacy through active learning
\citep[e.g.,][]{sauer2023active}. It helps to have a relatively higher
concentration of training data in regimes where dynamics are more challenging
to model, or across ``boundaries'' when changes are abrupt.

\section{Divide-and-conquer}

Divide-and-conquer GPs first partition the input space into
$k$ disjoint regions, i.e., $\mathbb{R}^d = \cup_{i=1}^k P_i$
with $\cap_{i=1}^k P_i = \emptyset$, and deploy (usually independent)
stationary GPs on each element of the partition.  
Let $n_i \subset 1, \dots, n$ denote the indices of the training 
data that fall in partition $P_i$, such that $X_{n_i} \in P_i$.
A partitioned zero-mean GP prior is then
$$
\mathbf{y}_{n_i} \stackrel{\mathrm{ind}}{\sim} \mathcal{N}\left(0,
	\boldsymbol{\Sigma}^{(i)}(X_{n_i})\right)
	\quad\textrm{for}\quad i = 1,\dots, k.
$$
Predictive locations are categorized into the existing partitions
$X_{p_i} \in P_i$, with independent posterior predictions 
following Eq.~(\ref{eq:gppred}), but with subscripts
$p \rightarrow p_i$ and $n \rightarrow n_i$ for each $i = 1,\dots,k$.
Each GP component uses a stationary covariance kernel $\boldsymbol{\Sigma}^{(i)}(\cdot)$
following Eq.~(\ref{eq:kernel}) or variations thereof.  Here, the 
superscript $(i)$ denotes the fact that the kernels on each partition 
may be parameterized disparately, or have different values for kernel
hyperparameters -- this is the key to driving nonstationary flexibility.
The pivotal modeling decision then becomes the choice of  
partitioning scheme, creating a patchwork of fits in the input space.

Motivated by ML applications to regression problems,
\cite{rasmussen2001infinite} first deployed divide-and-conquer GPs with the
divisions defined by a Dirichlet process (DP).  Unknown quantities, including
group assignments, DP concentration parameter, and covariance hyperparameters
for each GP component, were inferred using MCMC in a Gibbs framework. This
``infinite mixture'' of GPs strains tractability except for the smallest
problems.  The DP is perhaps too flexible, and may be the reason why the
empirical comparisons of this work were limited to a one-dimensional
illustrative example.  Focusing on spatial applications in two dimensions,
\cite{kim2005analyzing} later proposed a ``piecewise GP'' where the partitions
are defined from a Voronoi tesselation.  Again, Bayesian MCMC over the
partitions (tesselations in this case) was required.  While the Voronoi
tesselations with curved edges provide a flexible partitioning scheme, they
present a challenge to extrapolate into higher dimensions.  However, there
has been some good recent work toward that end \citep{pope2021gaussian}.

Expanding to the moderate dimensions common in surrogate modeling applications
warrants some reigning-in of partition flexibility. \cite{gramacy2008bayesian}
proposed a {\it treed Gaussian process} (TGP) which partitions the input space
using regression trees \citep{chipman1998bayesian}.  Partitions are
accomplished through a greedy decision tree algorithm, with recursive
axis-aligned cuts. Independent GPs are then fit on each partition (also known as
leaf nodes). The restriction of partitions to axis-aligned cuts is
parsimonious enough to allow for estimation in higher dimensional spaces.
TGPs naturally perform well when nonstationarity manifests along axis-aligned
partitions, yet learning the locations of the optimal partitions still
presents an inferential challenge.   
\citeauthor{gramacy2008bayesian} prioritize
UQ by performing MCMC sampling of the tree partitions, thus averaging over
uncertainty in the partition locations.  TGP implementation is neatly wrapped
in the {\sf R}-package {\tt tgp} on CRAN \citep{tgp}. 

\begin{figure}[ht!]
\centering
\begin{minipage}{5.5cm}
\includegraphics[scale=0.55, trim = 0 0 0 25]{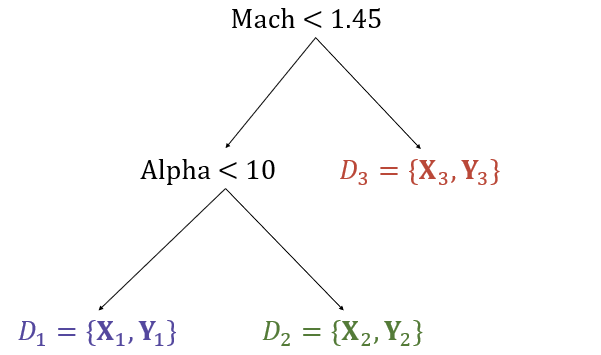}
\end{minipage}\hspace{2cm}
\begin{minipage}{6.5cm}
\includegraphics[scale=0.25, trim = 0 0 0 0]{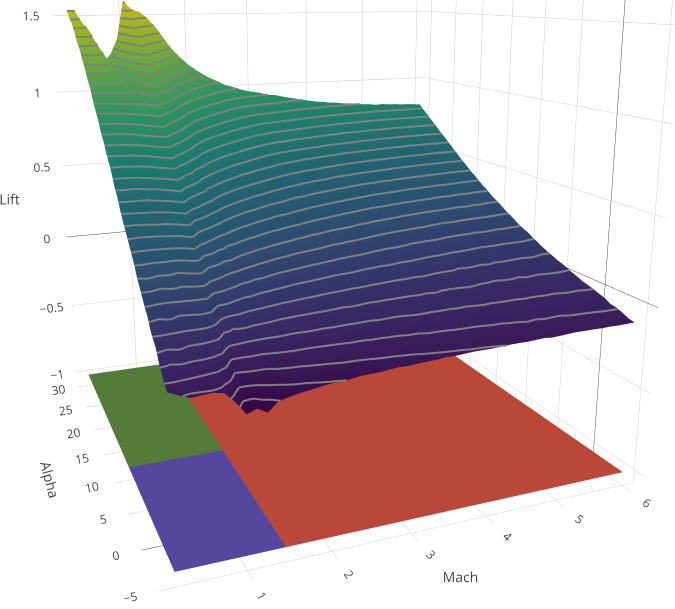}
\end{minipage}
\begin{minipage}{1cm}
\includegraphics[scale=0.33, trim = 0 0 0 20]{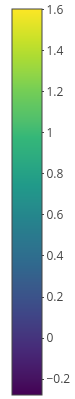}
\vspace{0.25cm}
\end{minipage}
\caption{{\em Left:} Example of a tree generated by TGP on an LGBB
dataset. Splits are made on speed (mach) 
and angle of attack (alpha). {\em Right:} Predictive surface plot 
of the rocket's lift. The partition on the input space generated 
by the tree is shown on the bottom panel, with colors corresponding 
to respective leaf nodes.}
\label{fig:tree}
\end{figure}

A motivating example that showcases the benefits of the TGP model is the
Langley Glide-Back Booster simulation \citep[LGBB;][Chapter
2]{pamadi2004aerodynamic,gramacy2020surrogates}. The simulator emulates the
dynamics of a rocket booster gliding through low Earth orbit. Inputs include
the rocket's speed (mach), angle of attack (alpha), and slip-side angle
(beta); one of the outputs of interest is the resulting lift force. The
dynamics of the simulator are known to drastically vary across mach values,
specifically when speeds cross the sound barrier. TGP model-fitting entertains
several different partitions of the three-dimensional input space. One of the
more commonly generated trees is shown in the left panel of Figure
\ref{fig:tree}, which heirarchically splits based on mach values of 1.45 and
alpha values of 10. The predictive surface fit using the TGP model, averaging
over all high-probability trees, is visualized in the surface plot on the
right, where testing locations are fixed at a beta value of 1. The partition
produced by the tree is illustrated underneath, on the $x$-$y$ plane; each
color-coded region corresponds to one of the three leaf nodes contained in the
tree on the left.  The model appropriately partitions the input space where
the dynamics of the response surface shift.  It first chooses a split at the
``divot'' observed at speeds under 1.45, then decides to further divide that
region based on smaller attack angles. \citet{bitzer2023hierarchical} recently
expanded upon the TGP model by allowing for partitions based on hyperplanes
which need not be axis-aligned (consider, for example, parallelograms instead
of rectangles in the bottom plane of the right panel).

So far, the traditional divide-and-conquer GPs that we've discussed have
focused on training data, via their inputs, as the ``work'' that is to be
divided.  What if we instead focused on the predictive locations as the avenue
of division?  After all, the overarching objective of the surrogate model is
to provide predictions (and UQ) at unobserved input configurations $X_p$.  We
can divide this ``work'' into $n_p$ jobs: predicting at each $\mathbf{x}_k \in
X_p$ for $k = 1, \dots, n_p$. Perhaps the earliest inspiration of this
approach stems from ``moving window'' kriging methods in which kernel
hyperparameters were estimated separately for each observation based on a
local neighborhood \citep{haas1990kriging}, generalizing the so-called
``ordinary kriging'' approach to geo-spatial modeling \citep{Matheron1971}.
Treating predictions as independent tasks sacrifices the ability to estimate
the entire predictive covariance, $\boldsymbol{\Sigma}^\star$ in
Eq.~(\ref{eq:gppred}), but point-wise variance estimates are sufficient for
most downstream surrogate modeling tasks.

Local approximate GPs \citep[laGP;][]{gramacy2015local} combine independent GP
predictions with strategic selection of conditioning sets. Rather than
conditioning on the entire training data $\{X_n, \mathbf{y}_n\}$ in
Eq.~(\ref{eq:gppred}), predictions for each $\mathbf{x}_k \in X_p$ condition
on a strategically chosen subset $\{X_{n_k},\mathbf{y}_{n_k}\}$.
Nonstationary flexibility comes from the independent nature of each GP; each
$\mathbf{x}_{k}$ may have its own kernel hyperparameterization, thus allowing
for the modeling of different dynamics at different locations. The original
motivation for this work was to circumvent the cubic computational bottlenecks
of GP inference that accompany large $n$ (by setting a maximum conditioning
set size $|n_k|$), but the nonstationary flexibility has been a welcome
byproduct \citep{sun2019emulating}.  Just as partition GPs rely on a choice of
partitioning scheme, local GPs rely on the choice of conditioning sets
\citep{emery2009kriging}.  The rudimentary approach is to select the training
data observations that are closest to $\mathbf{x}_k$ in Euclidean distance to
populate $X_{n_k}$ -- these are termed the ``nearest neighbors''.  In their
seminal work, \citeauthor{gramacy2015local} proposed strategic sequential
selection of points in each conditioning set based on variance reduction
criteria.  The {\sf R} package {\tt laGP} offers a convenient wrapper for
these local approximations \citep{laGP}.

\begin{figure}[ht!]
\centering
\begin{minipage}{7cm}
\includegraphics[scale=0.4,trim=0 0 0 20]{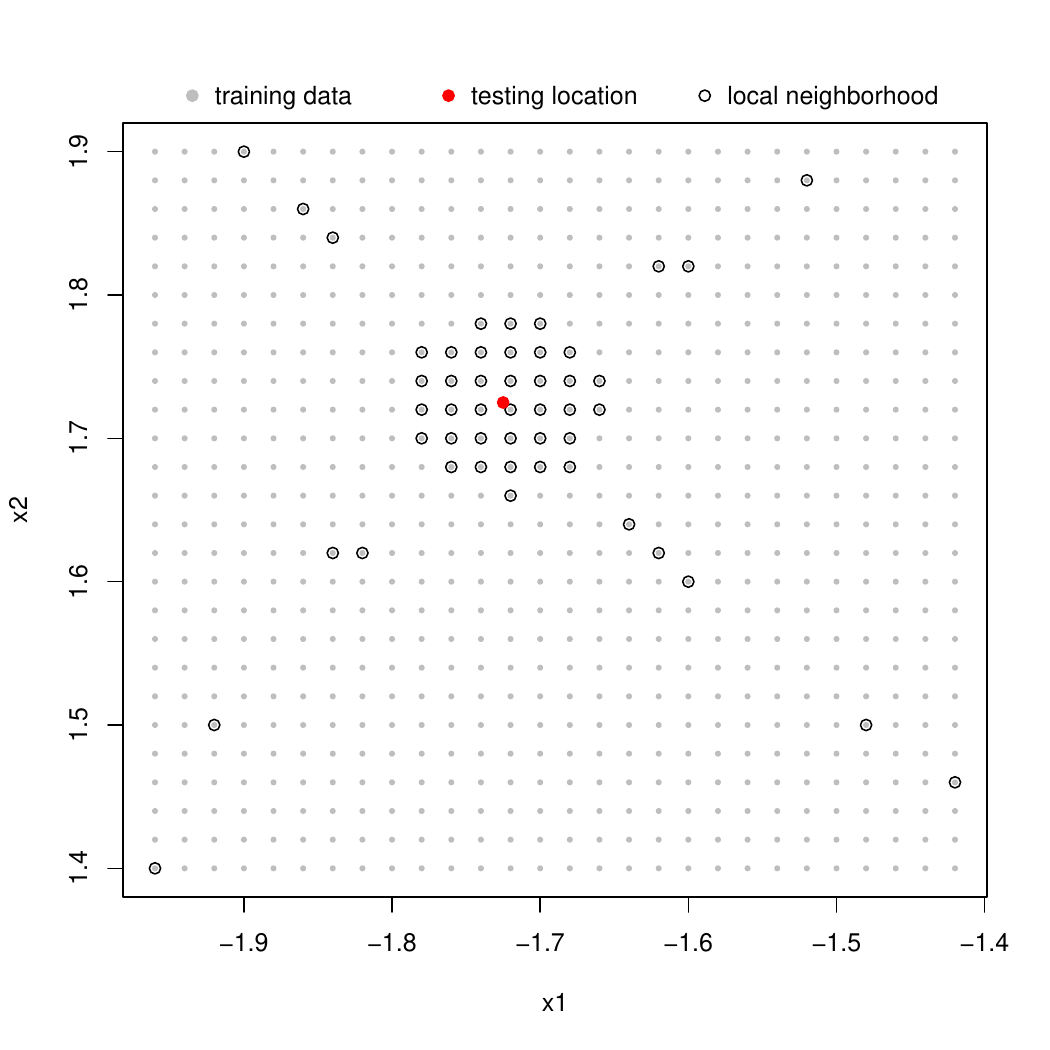}
\end{minipage} \hspace{0.5cm}
\begin{minipage}{7cm}
\includegraphics[scale=0.22,trim=0 0 20 50]{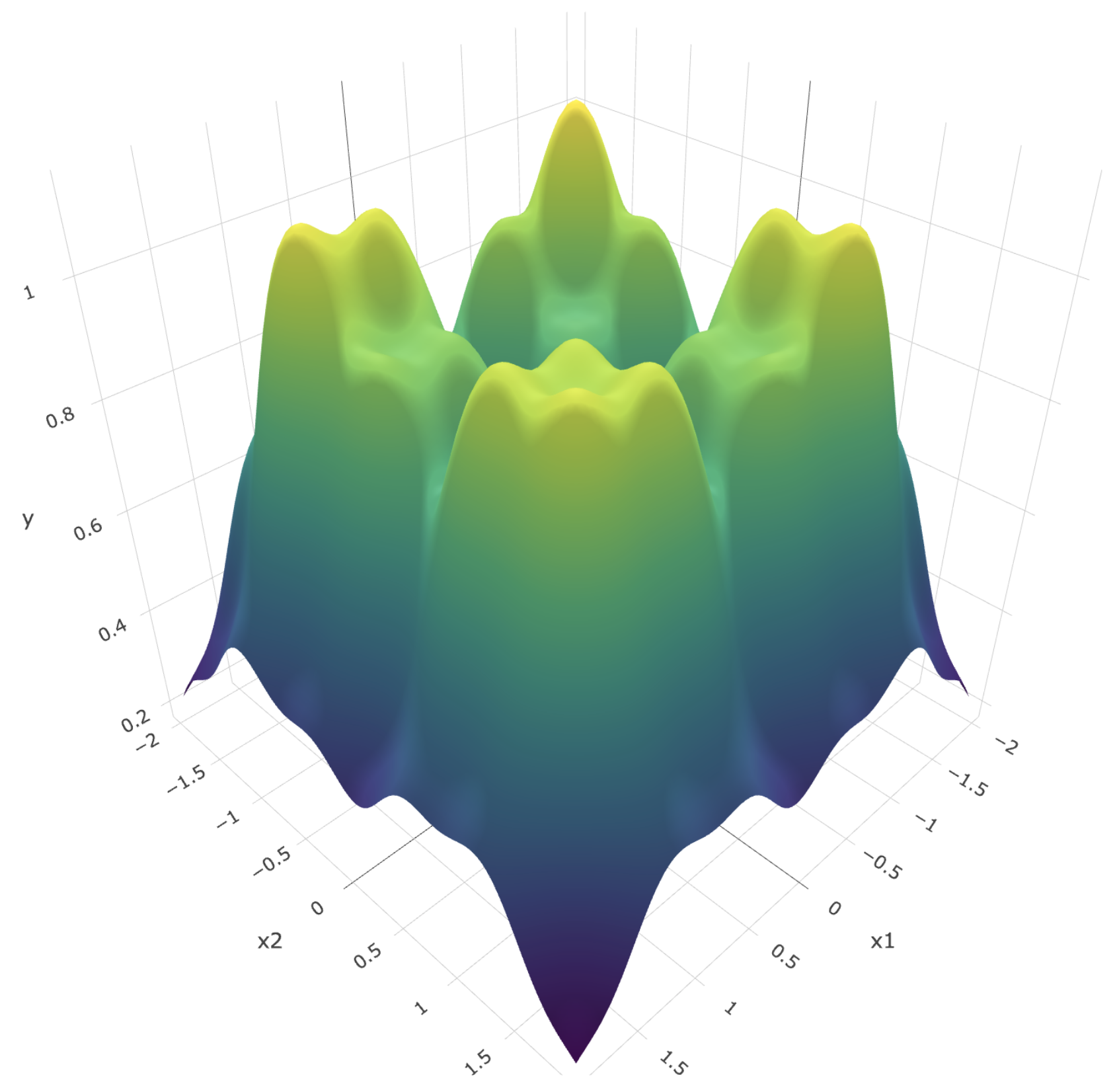} 
\end{minipage}
\begin{minipage}{1cm}
\includegraphics[scale=0.3]{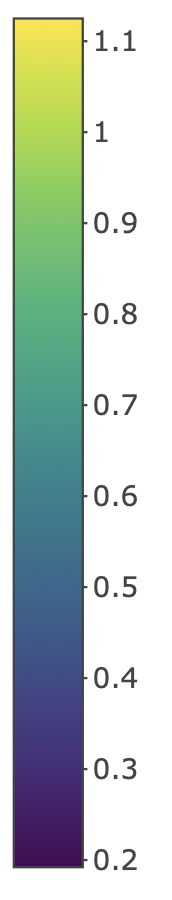} 
\end{minipage}
\caption{{\em Left:} Example of a local neighborhood (open circles), selected
from a large training set (grey dots) for predicting at the red dot.  {\em
Right:} A global fit based on a patchwork {\tt laGP}s applied
over a dense testing set.}
\label{fig:lagp}
\end{figure}

By dividing one big problem into many small ones, {\tt laGP} avoids the cubic
computational bottlenecks implicit in GP regression and creates an embarassingly
parallel computational task that is amenable to multi-core and
distributed computing \citep{gramacy2014massively}.  The left panel of Figure
\ref{fig:lagp} shows an example local neighborhood of size $n=50$
for predicting at the red dot.  Each of the grey dots is one of $N > 40K$
training data locations in $[-2,2]^2$. It goes without saying that you
couldn't fit an ordinary GP on a training data set that large, but you can use
a local approximation over a vast testing grid to estimate a complex, 
nonstationary surface, as shown in the right panel of the figure.  On an
eight-core machine, the fit and prediction steps take less than minute.

Divide-and-conquer GPs bring nonstationary flexibility by relying on
partitioned or local application of typical stationary GPs.  Yet this
flexibility comes at the expense of globality.  The independent nature of the
component GPs hampers the ability to learn any global trends. As one such
example, turn back to the Higdon function of Figure \ref{fig:higdon}. [We
omitted {\tt laGP} from this exercise since local GPs are focused on large
data sizes.] We presumed the training data was observed with an unknown noise
variance, but assumed that the variance was constant across the entire space
(aside from the {\tt hetGP} model which was just for demonstration). It would
be beneficial for a surrogate model to leverage all of the training data
simultaneously, learning from the linear region that the data is noise-free
and using that information to inform the interpolation of points in the left
``wiggly'' region.  Yet divide-and-conquer GPs are instead tasked with
estimating the noise independently on each division. The TGP fit in the lower
left panel provides a clear visual; it created a partition in the middle of
the space, fit separate GPs to both sides, but was unable to leverage
information between the two sides.  The search for a GP adaptation that is
both global and nonstationary leads us to our final class of models -- those
that utilize spatial warpings.

\section{Spatial warpings}\label{sec:warp}

Rather than adjusting the kernel or applying a multitude of 
independent GPs, ``warped'' GPs attempt to apply a regular
stationary GP to a new input altogether.  If the response is nonstationary
over $X$-space, perhaps we can find a new space (let's call it $W$-space)
over which the response is plausibly stationary.  Then a traditional GP
prior from $W$ to $\mathbf{y}$, i.e., $\mathbf{y} 
\sim N\left(\boldsymbol\mu, \boldsymbol{\Sigma}(W)\right)$
with stationary $\boldsymbol{\Sigma}(\cdot)$, would be an appropriate fit.
We consider $W$ as a ``warped'' version of $X$, hence
the name of this section;  it is the driver of nonstationary flexibility.
The characteristics of $W$ are a crucial modeling choice.  Ideally the
warping should be learned from training data (extensions that incorporate
expert domain specific information are of interest, but not yet thoroughly
explored), but finding an optimal and/or effective warping is a 
tall order.

Perhaps the earliest attempt to apply GPs on reformed inputs
was from \cite{sampson1992nonparametric}, who utilized ``spatial dispersion''
of the response values as the warping component.  
%They expressed spatial covariance over the spatial dispersions 
%rather than relative input distances,
%i.e., $\boldsymbol{\Sigma}_\mathrm{ns}(\mathbf{x}_i, \mathbf{x}_j) = \text{Cov}(y_i - y_j)$.  
Spatial dispersions ($W$ in our notation) were learned through a 
combination of multi-dimensional
scaling and spline interpolation, allowing for nonlinear warpings but not
accounting for uncertainty in the learned warpings.  \cite{schmidt2003bayesian} 
expanded upon this work by placing a GP prior over the warping, effectively
creating the hierarchical model
\begin{equation}\label{eq:dgp}
\begin{aligned}
\mathbf{y}_n \mid W &\sim N\left(\boldsymbol\mu_w, \boldsymbol{\Sigma}_w(W)\right) \\
W &\sim N\left(\boldsymbol\mu_x, \boldsymbol{\Sigma}_x(X_n)\right).
\end{aligned}
\end{equation}
This GP prior allowed for flexible nonlinear warpings and provided a natural
avenue for uncertainty quantification surrounding $W$. They utilized
Metropolis Hastings (MH) sampling of the unknown/latent $W$, wrapped in a
Gibbs framework with various kernel hyperparameters.
\citeauthor{schmidt2003bayesian} thus created the first deep Gaussian process, 
although they did not name it as such. DGPs did not gain popularity
until ten years later when those in the ML community reinvented it and
coined the DGP name \citep{damianou2013deep}, by analogy to DNNs -- more on
this parallel momentarily. There are several ways to compose a DGP, but the
simplest is to link multiple GPs through functional compositions (\ref{eq:dgp}).
This composition may be repeated to form
deeper models.  Intermediate layers remain unobserved/latent and must be
inferred.  \cite{dunlop2018how} have shown that this same model may be
formulated through kernel convolutions, and is thus a special subset of the
nonstationary kernel methods we discussed in Section \ref{sec:kernels}.

There are several key distinctions between the original work of 
\citeauthor{schmidt2003bayesian} and the DGPs that ML embraced 
a decade later.  First, \citeauthor{schmidt2003bayesian} focused on
two-dimensional inputs, meaning their warping $W$ was a matrix of 
dimension $n\times 2$.  They utilized a matrix Normal distribution
over $W$ (our representation in Eq.~(\ref{eq:dgp}) was strategically 
simplified), and in this smaller setting they were able to sample 
from the posterior distribution of the entire $W$ matrix using MH 
based schemes.  This proved too much of a computational 
burden for expanding to higher dimensions.  \cite{damianou2013deep}
proposed the simplication that each column of $W$ be conditionally
independent, leaving the two-layer DGP prior as
\[
\begin{aligned}
\mathbf{y}_n \mid W &\sim N\left(\boldsymbol\mu_w, \boldsymbol{\Sigma}_w(W)\right) \\
\mathbf{w}_i &\stackrel{\mathrm{ind}}{\sim} N\left(\boldsymbol\mu_x, \boldsymbol{\Sigma}_x(X_n)\right),
	\;\; i = 1,\dots, p 
\end{aligned}
\quad\textrm{where}\quad
W = \begin{bmatrix} \mathbf{w}_1 & \mathbf{w}_2 & \dots 
	& \mathbf{w}_p \end{bmatrix}.
\]
These columns are often referred to as ``nodes.''  The number of nodes
is flexible, but most commonly set to match the input dimension, i.e., $p = d$.
Now the parallel between DGPs and DNNs is clearer: a DGP is simply a DNN
where the ``activation functions'' are Gaussian processes.  Posterior inference
requires integrating out the unknown latent warping,
\begin{equation}\label{eq:dgp_post}
\mathcal{L}(\mathbf{y}\mid X) = \int\dots\int \mathcal{L}(\mathbf{y}\mid W)
	\prod_{i=1}^p \mathcal{L}(\mathbf{w}_i \mid X) \;\; 
	d\mathbf{w}_1,\dots,d\mathbf{w}_p,
\end{equation}
which is not tractible in general.  Extensions to deeper models require
additional compositions of GPs, with even more unknown functional quantities 
to integrate 
\citep[see][for thorough treatment of a three-layer DGP]{sauer2023active}.  
Faced with this integral, many in ML embrace approximate
variational inference (VI) in which the posterior of Eq.~(\ref{eq:dgp_post})
is equated to the most likely distribution from a known target family
\citep[e.g.,][]{damianou2013deep,bui2016deep,salimbeni2017doubly}.
This approach replaces integration with optimization, but in so 
doing it oversimplifies UQ.  Additionally, it is unable to address the
multi-modal nature of the posterior that is common in DGPs 
\citep{havasi2018inference}.

\begin{figure}[ht!]
\vspace*{0.5cm}
\centering
\begin{minipage}{7cm}
\includegraphics[scale=0.35,trim=0 10 20 0]{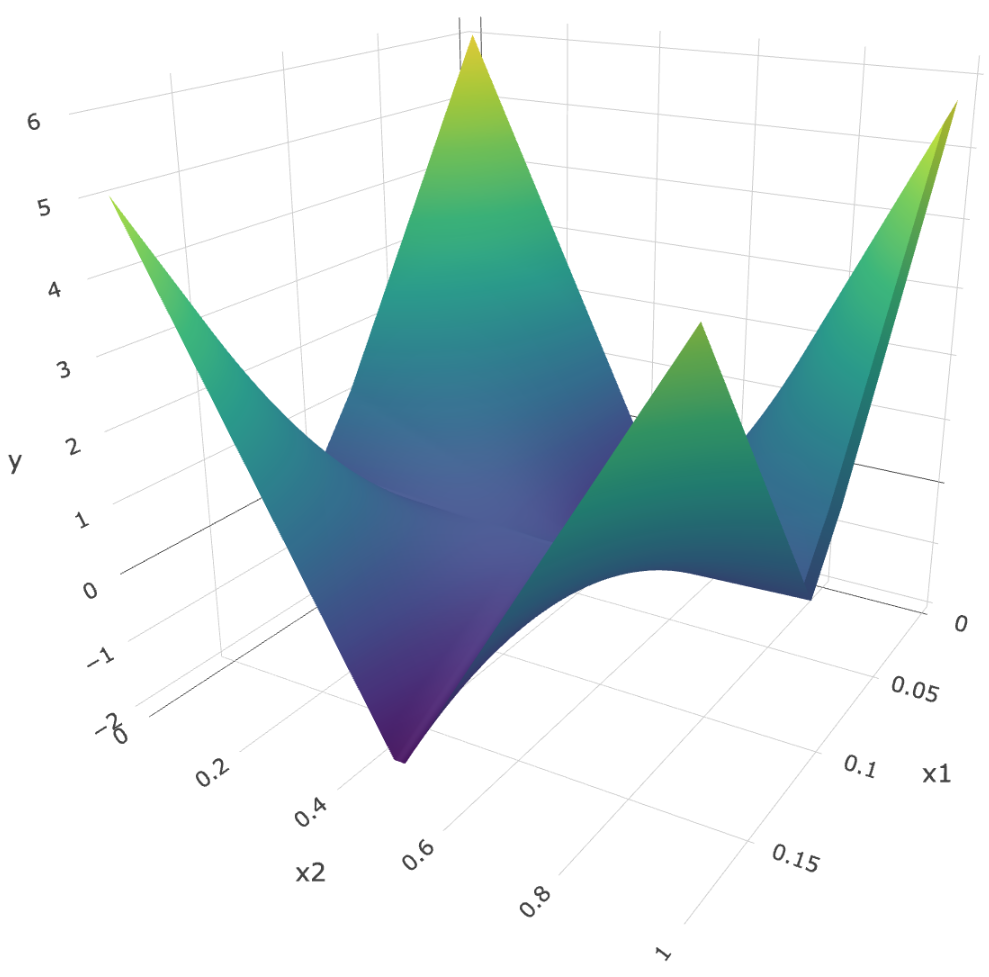} 
\end{minipage}
\begin{minipage}{1cm}
\includegraphics[scale=0.4,trim=50 0 50 0]{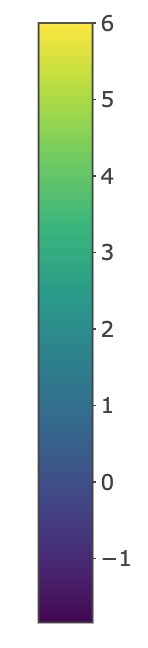} 
\end{minipage}
\begin{minipage}{6cm}
\includegraphics[scale=0.4,trim=20 20 0 0]{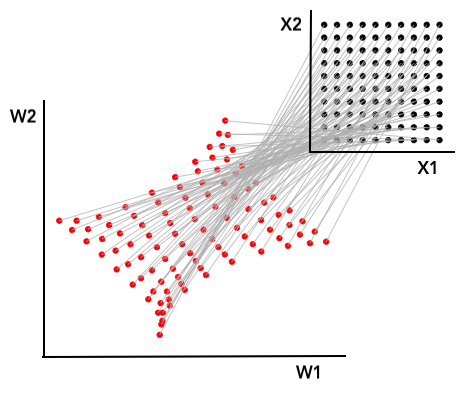}
\end{minipage}
\caption{{\em Left:} The two-dimensional G-function.
{\em Right:} A posterior ESS sample of latent layer $W$ from a two-layer DGP
fit to training data from the G-function. Together, ``nodes'' of 
$W$ act as a warped verson of $X$, allowing the DGP to accommodate 
the steep inclines along the diagonals.}
\label{fig:dgp}
\end{figure}

Surrogate modeling tasks demand broader uncertainty quantification. Full
posterior integration through MCMC sampling of the latent warping offers a
solution, but Metropolis Hastings type samplers have been shown to suffer from
high rejection rates and poor mixing in large DGP setups. To combat this,
\cite{sauer2023active} proposed a fully-Bayesian inferential scheme for DGPs
that employs elliptical slice sampling \citep[ESS;][]{murray2010elliptical} of
latent Gaussian layers.  The rejection-free proposals of the ESS algorithm
work well in DGP settings and are able to explore multiple modes more readily
than MH counterparts.   A similar method by \citet{ming2023deep} was proposed
around the same time, using ESS for the latent warping layer but taking a more
thrifty approach to hyperparameter inference. \cite{barnett2024monotonic} offered 
an alternative ESS implementation which can force DGP latent warpings to be 
monotonic, but it requires each input dimension to be individually warped.
Others have embraced Hamiltonian Monte Carlo sampling as an alternative 
to ESS with similar outcomes \citep{havasi2018inference}. 
Full propagation of uncertainty through posterior
samples of $W$ can be crucial for downstream surrogate modeling tasks
including active learning \citep{sauer2023active}, Bayesian optimization
\citep{gramacy2022triangulation}, and reliability analysis \citep{booth2024hybrid}.

The fully-Bayesian MCMC implementation of DGPs is wrapped in the {\tt deepgp}
{\sf R}-package on CRAN \citep{deepgp}.  We provide two visuals of DGP
warping/flexibility utilizing this package.  The first is shown in the lower
center/right panels of Figure \ref{fig:higdon}.  The grey lines display ESS
samples of latent $W$, fit to the training data from the Higdon function.  The
steep slope in the left region has the effect of ``stretching'' the inputs
where there is high signal.  The flattening-off of the samples in the right
region effectively ``compresses'' inputs in the linear region. Notice how ESS
samples of $W$ bounce back-and-forth between modes (positive slopes and
negative slopes); since only pair-wise distances feature in the stationary
kernel of the outer layer, these mirror images are equivalent. The resulting
DGP fit is shown in the lower center panel; the warped model is able to
accommodate the two piecewise regimes while leveraging global learning of
kernel hyperparameters. The second visual concerns the two-dimensional
G-function \citep{marrel2009calculations}, displayed in the left panel of
Figure \ref{fig:dgp}.  This function is characterized by stiff peaks and steep
valleys.  In two dimensions, latent $W$ is comprised of two nodes, each a
conditionally independent GP over the inputs.  The right panel of Figure
\ref{fig:dgp} visualizes a single ESS sample of $W$, interpreted as a warping
from evenly gridded $X$.  The response surface is plausibly stationary over
this warped regime, thus allowing for superior nonstationary fits.
The {\tt dgpsi} package \citep{dgpsi} provides a similar implementation of DGPs
with ESS in {\sf python}.

DGPs offer both global modeling and nonstationary flexibility, but they come
with a large computational price tag.  Posterior sampling requires thousands
of samples, usually wrapped in a Gibbs scheme with many Gaussian likelihood
evaluations required for each iteration.  For training data sizes above
several hundred, these computations become prohibitive.  The machine learning
approach -- approximate inference by VI -- partially addresses this issue by
turning inference by posterior marginalization (i.e., by integral) into
inference by optimization, but such maneuvers run the risk of undercutting UQ.
Moreover, optimization of the Kullback-Leibler divergence between the desired
DGP posterior and the specified target family requires thousands of
iterations, on par with the iterations needed for MCMC mixing. Our experience
in the context of DGPs is that the speedups offered by VI are marginal at
best, modulo myriad other Monte Carlo considerations such as mini-batch sizes
and MCMC iterations.  Big computational gains require further approximation.
The predominant tool here is inducing points
\citep[IPs;][]{snelson2006sparse}; all of the VI references we have mentioned
so far, as well as the Hamiltonian Monte Carlo sampling implementation of
\cite{havasi2018inference}, make use of IP approximations. But IPs come with
several pitfalls: without large quantities and/or optimal placement, they
provide blurry low-fidelity predictions \citep{wu2022variational}. Some have
embraced random feature expansions as alternatives to inducing points
\citep{marmin2022deep}, but perhaps the most successful alternative has been
Vecchia approximation \citep{vecchia1988estimation,sauer2023vecchia}.  The
Vecchia approximation forms the basis of scalability in the {\tt deepgp}
package and is also offered in {\tt dgpsi}.  In our own experience, the
Vecchia-based DGP approach is more robust and user friendly -- offering more
accurate predictions with better UQ out of the box -- compared to VI/IP
alternatives, and others in similar spirit such as the Python libraries {\tt
GPflux} \citep{gpflux} and {\tt GPyTorch} \citep{gpytorch}. Occasionally we
can fine-tune these libraries to get competitive results in terms of accuracy,
but UQ still suffers distinctly. It is worth noting that most software for
Vecchia-approximated (D)GPs offer a ``lite'' prediction option, which is akin
to the nearest neighbor conditioning of {\tt laGP}.  In these settings, the
key distinction is the globality of the Vecchia-approximated covariance used
for model training.

\section{Empirical benchmarking}

Here we present a comparison of a selection of the aforementioned
nonstationary GP surrogates on a real-world computer simulation.  The {\it
Test Particle Monte Carlo} simulator was developed by researchers at Los
Alamos National Laboratory to simulate the movement of satellites in low earth
orbit \citep{mehta2014modeling}.  The simulation is contingent on the
specification of a satellite mesh.  \citet{sauer2023vecchia} entertained
nonstationary DGP surrogates on the GRACE satellite; we instead consider the
Hubble space telescope, which has an additional input parameter, for a total
of 8 dimensions.  The response variable is the amount of atmospheric drag
experienced by the satellite.  See \citet[][Chapter 2]{gramacy2020surrogates}
for further narration of this simulation suite.

We utilize a data set of 1 million simulation runs provided by
\citet{sun2019emulating}, selecting disjoint random subsets of size $n = n_p =
10{,}000$ for training and testing.  Although previous studies have looked at
larger training/testing sets, we chose a more moderate one to entertain a
wider set of competitors.  We entertain the following surrogate models:
\begin{itemize}
	\item GP SVEC: stationary GP utilizing scaled-Vecchia approximation 
	\citep{katzfuss2020scaled}.
	\item laGP: local approximate GP \citep{gramacy2015local} using the {\tt laGP}
	package \citep{laGP}.
	\item TGP: treed-GP \citep{gramacy2008bayesian} using the {\tt tgp} package
	\citep{tgp}.
	\item DGP ESS: Bayesian DGP with elliptical slice sampling and
	Vecchia approximation \citep{sauer2023active,sauer2023vecchia} using the 
	{\tt deepgp} package \citep{deepgp}.
	\item DGP DSVI: DGP with approximate ``doubly stochastic'' variational 
	inference and inducing point approximations \citep{salimbeni2017doubly} using the
	{\tt GPflux} package \citep{gpflux}.
\end{itemize}
Any larger $n$ would preclude including TGP.  A moderate
$n$ has the added benefit of accentuating performance disparities between
stationary and nonstationary models.  
%Much larger $n$ complicates computation,
%but dampens discrepancies between deep and ordinary GPS, as both are universal
%approximators in a certain sense.  

\begin{figure}[h]
\centering
\includegraphics[width=16cm,trim=0 0 0 20]{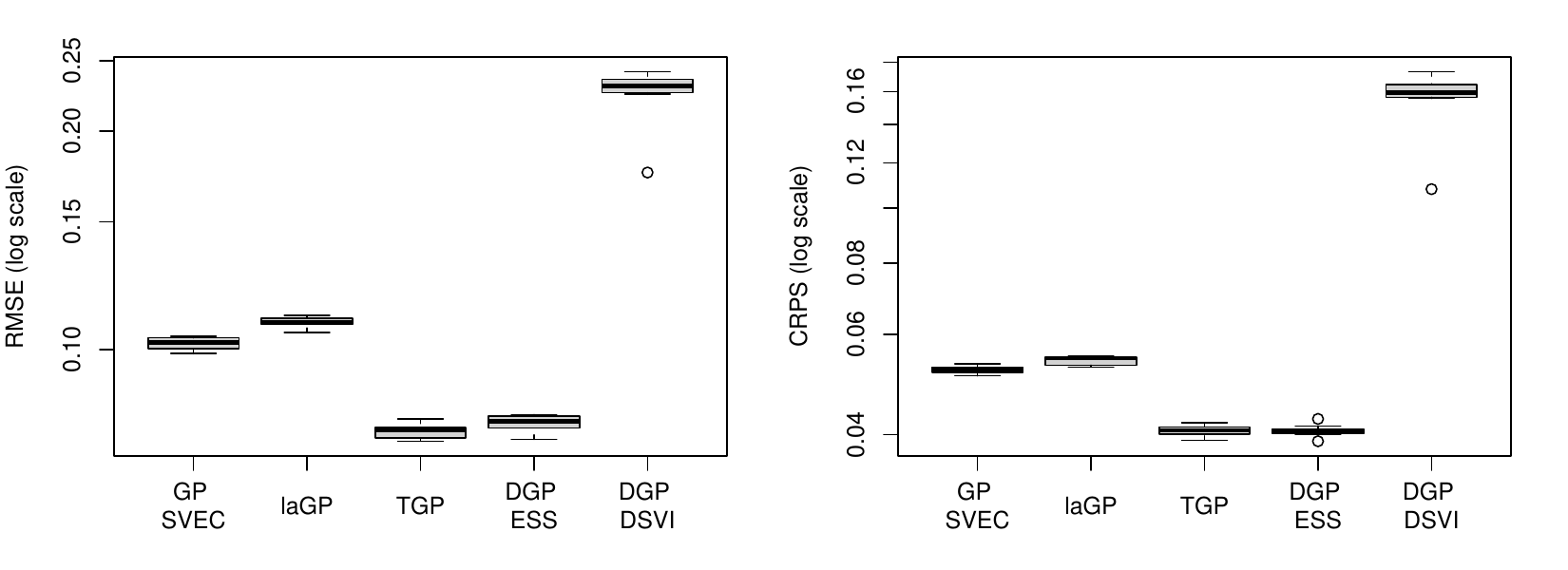}
\caption{Surrogate performance on the 8-dimensional Hubble Space Telescope 
simulation across 10 MC repetitions.}
\label{fig:satdrag}
\end{figure}

We follow \citet{sun2019emulating} in fixing the noise level at $\nu = 1\times
10^{-4}$.  Separable lengthscales estimated from the GP SVEC model are used to
pre-scale inputs prior to fitting the other surrogate models [except for TGP
 since the {\tt tgp} package conducts its own scaling]. Input
pre-scaling and similar analogues have been shown to improve surrogate
performance and have become standard
\citep[e.g.,][]{sun2019emulating,katzfuss2020scaled,
wycoff2021sensitivity,kang2023correlation}.  Model performance is reported by
root mean squared error (RMSE, lower is better) and continuous rank
probability score \citep[CRPS;][Eq.~20, negated so lower is
better]{gneiting2007strictly}.  While RMSE captures surrogate predictive
accuracy, CRPS incorporates posterior UQ. Results across 10 Monte Carlo
repetitions, with re-randomized training and testing sets, are shown in Figure
\ref{fig:satdrag}.  Reproducible code for this experiment is provided in our
public git repository.\footnote{{\tt
https://bitbucket.org/gramacylab/deepgp-ex/}}

Of the five methods, DSVI's approach to DGP inference stands out as being
particularly poor. We don't have a good explanation for that except that the
class of problems it was engineered for -- low-signal large-data regression
and classification for machine learning tasks -- is different than our
computer modeling context, with modest training data size and high
signal-to-noise ratios.  VI, by replacing an integral with optimization,
undercuts on UQ, and we suspect the IPs sacrifice too much fidelity in this
instance.  Observe that {\tt laGP} performs relatively well, but interestingly
not as well as an ordinary GP (represented by GP SVEC). Our explanation here
is that {\tt laGP} was designed for massive data settings on the scale of
millions of observations.  It was developed primarily with speed and
parallelization in mind, with nonstationary flexibility being a byproduct of
its divide-and-conquer approach.  These data do indeed benefit from deliberate
nonstationary modeling, as indicated by the TGP and DGP ESS comparators. We
believe TGP edges out DGP on this example because the process benefits from
crude, axis-aligned partitioning. %The additional flexibility
%input-warping represents an hazard, by comparison, though not one with a big
%downside. 
We observed a high degree of TGP partitioning on the eighth,
``panel-angle'' input.  It would seem the orientation of Hubble's solar
panels is driving regime changes in drag dynamics.  Both TGP and DGP ESS
utilize a fully Bayesian approach to inference for all unknown quantities.
Consequently they have high predictive accuracy (via RMSE) and UQ (via CRPS).

\section{Discussion}

Nonstationary GP modeling is a rapidly developing research area.  For many
computer experiments we are now presented with the opportunity to learn
complex higher fidelity dynamics, which may or may not be present in the
system under study. Residual diagnostics can provide insight into
insufficiences in stationary GPs \citep{bastos2009diagnostics}, but with a
growing array of possible nonstationary models, choosing an effective
surrogate is increasingly daunting. One attractive aspect of many of the best
nonstationary surrogates is that they nest simpler models as special cases.
For example, a DGP nests an ordinary GP through an identity warping. TGP
nests a GP when the posterior prefers not to split.  A heteroskedastic GP
nests an ordinary GP when the variance is constant.  This structure reduces
the ``choice'' amongst potential surrogates of varying nonstationary
complexity to the ``problem'' of model selection, averaging, and
regularization, which are important aspects of any modeling enterprise. One
advantage of a Bayesian approach is that all three of those ``problems'' are
solved via the natural regularization that a prior provides when all
quantities are modeled via the posterior.  Nested models open the door to
traditional model selection procedures such as Bayes factors, Chi-squared
tests, and similar techniques \citep{jacquier2004bayesian}.  When overfitting
is a concern, the easiest recourse is a validation exercise: fit nonstationary
and stationary models and assess some predictive scoring rule out-of-sample,
like in our exercise of Figure \ref{fig:satdrag}.

\bibliographystyle{jasa}
\bibliography{gp_uq}

\end{document}